\documentclass[conference,letterpaper]{IEEEtran}

\usepackage{cite}
\usepackage{color}
\usepackage{enumerate}
\usepackage{booktabs}
\usepackage{setspace}
\usepackage{bm}
\usepackage{amsthm}
\usepackage[cmex10]{amsmath}
\usepackage{colortbl}
\usepackage{stfloats}
\usepackage{subfigure}
\usepackage{amssymb}

\usepackage[linesnumbered,ruled]{algorithm2e}

\usepackage{multirow}
\usepackage{xcolor}
\usepackage{graphicx}
\usepackage{geometry}
\usepackage{stmaryrd}
\usepackage{caption}
\usepackage{relsize}
\theoremstyle{definition}

\usepackage{epstopdf}
\IEEEoverridecommandlockouts
\begin{document}

%
\title{Emergency Localization for Mobile Ground Users: An Adaptive UAV Trajectory Planning Method }
\vspace{-0.6cm}
\author{\IEEEauthorblockN{Zhihao Zhu\IEEEauthorrefmark{1},
Jiafan He\IEEEauthorrefmark{2}, 
Luyang Hou\IEEEauthorrefmark{1}, 
Lianming Xu\IEEEauthorrefmark{3},
Wendi Zhu\IEEEauthorrefmark{1},
and Li Wang\IEEEauthorrefmark{1}
\vspace{-0.2cm}
}\\
\small 
\centerline{\IEEEauthorrefmark{1}School of Computer Science, Beijing University of Posts and Telecommunications, Beijing, China}\\
\IEEEauthorrefmark{2}Science and Technology on Information Systems Engineering Laboratory, Nanjing Research Institute of Electronics Engineering, Nanjing, China\\
\IEEEauthorrefmark{3}School of Electronic Engineering, Beijing University of Posts and Telecommunications, Beijing, China\\
Email: {\{zhuzhihao, luyang.hou, xulianming, zhuwendi, liwang\}@bupt.edu.cn}, {hejiafan\_cetc28@sina.com}\\
\vspace{-1cm}
\thanks{The work was supported by Key Laboratory of Information System Engineering, No. 05202206. (\textit{Corresponding author: Li Wang})}
}
\vspace{-0.5cm}
\maketitle

\begin{abstract}
In emergency search and rescue scenarios, the quick location of trapped people is essential. However, disasters can render the Global Positioning System (GPS) unusable. 
Unmanned aerial vehicles (UAVs) with localization devices can serve as mobile anchors due to their agility and high line-of-sight (LoS) probability. 
Nonetheless, the number of available UAVs during the initial stages of disaster relief is limited, and innovative methods are needed to quickly plan UAV trajectories to locate non-uniformly distributed dynamic targets while ensuring localization accuracy. 
To address this challenge, 
we design a single UAV localization method without hovering, use the maximum likelihood estimation (MLE) method to estimate the location of mobile users and define the upper bound of the localization error by considering users' movement.
Combining this localization method and localization error-index,
we utilize the enhanced particle swarm optimization (EPSO) algorithm and edge access strategy to develop a low complexity localization-oriented 
adaptive trajectory planning algorithm. 
Simulation results demonstrate that our method outperforms other baseline algorithms, enabling faster localization without compromising localization accuracy.
\end{abstract}

\begin{IEEEkeywords}
UAV, localization, MLE, trajectory planning, search and rescue
\end{IEEEkeywords}

\section{Introduction}

Emergency localization refers to the swift and precise location of people or assets in emergencies, such as rescue, search, and disaster relief operations. This type of localization is crucial for successful search and rescue efforts, as it allows rescue teams to quickly locate and extract trapped individuals.
However, natural disasters may damage cellular infrastructure and hinder communication \cite{9355403}, while dense forests in mountainous regions can undermine the effectiveness of GPS localization \cite{9355402}. 
In such circumstances, the use of alternative ranging technologies, such as received signal strength indicator (RSSI), proves advantageous due to its ease of use and low cost. 
However, the ranging results of RSSI will be affected by the random factors of environmental shadows, resulting in weak link problems.
For weak links, some literature introduces caching resources to enhance communication:
\cite{7558170} proposed a local search algorithm based on hypergraph, \cite{ 9126265} proposed a matching algorithm based on multi-target bilateral preference lists, \cite{8234659} proposed a distributed matching algorithm based on Roth and Vande Vate, but the above works cannot support RSSI localization. 
And UAVs can establish reliable communication links through LoS connections to solve the problem of weak links\cite{8377137}. 
Additionally, UAVs can be deployed on demand and provide flexible mobility \cite{8752017}, making them ideal for search and rescue operations.

Recently, there has been a growing interest in utilizing UAVs as mobile anchors for ground target localization.
Notably, SLMAT \cite{7887767} and $\mathbf\Sigma$-SACN \cite{8777149} have been proposed as UAV localization along fixed trajectories to locate uniformly densely distributed targets.
However, these methods are less applicable in emergency scenarios where trapped people often exhibit sparse and non-uniform distributions.
To address this limitation, a solution was introduced in \cite{8960453} that employs map discretization as the state space for UAVs and utilizes reinforcement learning (RL) for autonomous UAV trajectory localization of targets. However, map discretization sacrifices localization accuracy and introduces high algorithmic time complexity.
Additionally, a UAV non-hover strategy for ground target localization to save localization time was proposed in \cite{9739714}, but it did not consider the trade-off between localization accuracy and flight length. 
Importantly, the existing localization methods do not adequately account for the mobility of the target, thereby limiting localization accuracy. 
In contrast, studies \cite{9896671} and \cite{9247418} enhance the Kalman filter and the least squares method, respectively, by leveraging the Doppler effect to provide localization services for mobile users. However, these studies assume that the user will always remain within the coverage of the anchor and are not suitable for emergency search and rescue scenarios.

Given the current research landscape, there is an urgent need for innovative methods to plan flight trajectories for a limited number of UAVs oriented toward locating non-uniformly distributed dynamic targets in emergency search and rescue scenarios.
Addressing these challenges would contribute significantly to advancing the field of UAV-based ground target localization, particularly in search and rescue scenarios.
We divide the localization task into initial scan and accurate localization for dynamic trapped people localization in emergencies. The contributions of this article are:

\begin{itemize}
	\item[$\bullet$] The UAV non-hovering strategy is designed to reduce the task completion time, using MLE to estimate the location of the mobile trapped people through single anchor point localization. The upper limit of localization error is defined by considering trapped people's movement.
\end{itemize}
\begin{itemize}
	\item[$\bullet$] In the accurate localization stage, to reduce the algorithm's time complexity, we decompose the problem and propose the EPSO algorithm. Additionally, we integrate the edge access strategy into the algorithm for efficient adaptive trajectory planning.
\end{itemize}
\begin{itemize}
	\item[$\bullet$]  The simulation results show that compared with existing solutions, the proposed solution has a shorter task completion time and lower algorithm time complexity while ensuring localization accuracy.
\end{itemize}

\section{System Model And Problem Formulation}

\subsection{System Model}

We consider a search and rescue scenario where $M$ UAVs fly over a mountainous forest disaster area to search and locate the trapped people whose number and exact location are unknown. 
Each trapped person is equipped with a wireless communication device that periodically emits a probe request signal, while each UAV continuously collects RSSI measurements of different trapped people within its communication coverage range during the flight. 
Subsequently, based on the RSSI measurements collected at different locations, each UAV uses the ranging model \cite{7451182} to calculate the distance between itself and the trapped person and estimate the location of the trapped person.
All UAVs embark on their rescue mission from a designated starting point and return after completing the mission. 
The sets of UAVs and trapped people are denoted as $\mathcal{M}=\{1,...,M\}$ and $\mathcal{S}=\{1,...,S\}$, respectively. 

The location of $m$-th UAV in time slot $t$ is expressed as $\mathbf{q}_{m}(t)=(x_{m}(t),y_{m}(t),h), m\in\mathcal{M}$, where $0 \leq x_{m}(t) \leq L_x$, $0 \leq y_{m}(t) \leq L_y$, $h$ is a fixed flying altitude, 
$L_x$ and $L_y$ represent the disaster area's length and width, respectively.
Similarly, the location of the $s$-th trapped person in time slot $t$ is denoted as $\mathbf{w}_{s}(t)=(x_{s}(t),y_{s}(t),0) $, with $ 0 \leq x_{s}(t) \leq L_x$ and $0 \leq y_{s}(t) \leq L_y$.
Besides, let $\mathbf{v}_{m}(t)$ and $ \mathbf{v}_{s}(t)$ denote the velocity of $m$-th UAV and $s$-th trapped person in time slot $t$, respectively. We assume that the maximum velocity of each UAV and trapped person are $V_{max}$ and $v_{max}$, respectively.
Therefore, the location of $m$-th UAV and $s$-th trapped person in time slot $t+1$ can be expressed, respectively,
\begin{align}
	\label{eq_q_m_t}
	&\mathbf{q}_{m}(t+1) = \mathbf{q}_{m}(t) + \mathbf{v}_{m}(t), \\
	\label{eq_w_m_t}
	&\mathbf{w}_{s}(t+1) = \mathbf{w}_{s}(t) +  \mathbf{v}_{s}(t).
\end{align}
Therefore, the distance between $m$-th UAV and $s$-th trapped person in time slot $t$ is given by
\begin{align}
	\label{eq5}
	d_{m,s}(t)= 
	& \| \mathbf{q}_{m}(t)  - \mathbf{w}_{s}(t) \| .
\end{align}

With RSSI ranging model, the estimated distance $r_{m,s}(t)$ between the $m$-th UAV and the $s$-th trapped person for a given distance $d_{m,s}(t)$ in time slot $t$ can be expressed as follows in \cite{7451182}For T time slots from ts
m,s to te
m,s, generate T sets according
to
\begin{equation}
	r_{m,s}(t)=d_{m,s}(t)10^{-\frac{\Psi}{10\eta}}=d_{m,s}(t)\mathrm{\Psi_{LN}}^{-1/\eta},
	\label{eq1}
\end{equation}
where $\eta$ is the path loss coefficient that quantifies the signal power decays with distance $d_{m,s}(t)$ and $\Psi \sim {\mathcal{N}}(0,\sigma_\Psi^2)$. $\mathrm{\Psi_{\textnormal{LN}}}^{-1/\eta} = 10^{-\frac{\Psi}{10\eta}}$ follows log-normal distribution with base $e$ and parameters $\mu=0$, $\sigma^2 = \frac{\sigma_{\Psi}^2}{\xi^2\eta^2}$, $\xi=10\textnormal{log}(e)$. Thus,  $r_{m,s}(t) \sim {\textit{LogN}}(\textnormal{ln}d_{m,s}(t),\frac{\sigma^2}{\eta^2})$, of which the probability density function (PDF) can be expressed as 
{
    \small
	\begin{equation}
		\label{eq6}
		f(r_{m,s}(t)) = 
		\frac{1}{{r_{m,s}(t)\sqrt{2\pi} \sigma}{/\eta}} \exp \left(\frac{-(\ln ( r_{m,s}(t) / d_{m,s}(t) ) )^2}{2\sigma^2/\eta^2}\right),
	\end{equation}
}
and the statistical mean of the ranging error can be expressed as follows in \cite{7451182}
\begin{equation}
	\label{range_error_mean}
	m_{\xi}(d_{m,s}(t)) = d_{m,s}(t)(1- e^{\frac{\sigma^2}{2\eta^2}}).
\end{equation}

The communication energy consumption can be ignored compared with the others. The propulsion energy of UAV with velocity $V$ can be expressed as follows in \cite{7888557}
{
\begin{equation}
	E  =  P_0(1+ \frac{3V^2}{U^2}) + P_1 \sqrt{
		\sqrt{1+\frac{V^4}{4v_r^2} } - \frac{V^2}{2v_r^2} 
	} + \frac{1}{2}A V^3, \label{eq3}
\end{equation}
}where $P_0$ represents the power required by the rotor blades of the UAV due to air resistance when the UAV is hovering; $P_1$ represents the power required by the UAV rotor to generate lift when the UAV is hovering; $U$ represents the tip speed of the UAV rotor blade; $v_r$ represents the average rotor speed of the UAV in the hovering state; $A$ represents the parameters related to air resistance, body friction and landing gear resistance that the UAV is subjected to during flight.

\subsection{Problem Formulation}

We divide all the trapped people into $M$ non-overlapping groups, i.e. $\mathcal{G}_m \subseteq \mathcal{S}, 1\leq m\leq M, \mathcal{G}_{1} \cap \mathcal{G}_{2} = \varnothing, \cup_{m=1}^M \mathcal{G}_{m} = \mathcal{S}$, where $\mathcal{G}_m$ represents the set of trapped people that need to be located by $m$-th UAV. 
Denote $\mathcal{L}_m = \{\mathbf{p}_{m,0},\mathbf{p}_{m,1},...,\mathbf{p}_{m,|\mathcal{G}_m|},,\mathbf{p}_{m,|\mathcal{G}_m|+1} \} $ as the coordinate point set where $m$-th UAV locates the trapped people in $\mathcal{G}_m$ and the departure and arrival points of the $m$-th UAV performing the localization task.
Let ${\Gamma_m} = \{ \kappa_{m,0},\kappa_{m,1},...,\kappa_{m,|\mathcal{G}_m|},\kappa_{m,|\mathcal{G}_m|+1} \}$ be a permutation of coordinate point labels in $\mathcal{L}_m$, then ${\Gamma_m}$ represents a serving ordered of tour $( \mathbf{p}_{m,\kappa_{m,0}},\mathbf{p}_{m,\kappa_{m,1}},...,\mathbf{p}_{m,\kappa_{m,|\mathcal{G}_m|}},\mathbf{p}_{m,\kappa_{m,|\mathcal{G}_m|+1}} )$. Denote $T_m$ as the total time of $m$-th UAV for accomplishing the localization task. To complete the search and rescue mission as soon as possible, each UAV should fly along the line segment connecting $\mathbf{p}_{m,\kappa_{m,l}}$ and $\mathbf{p}_{m,\kappa_{m,l-1}}$ with the maximum velocity $V_{max}$, $0 \leq l \leq |\mathcal{G}_m|$, then we have
\begin{align}
	\label{eqtm}
	T_m = \frac{\sum_{l=1}^{|\mathcal{G}_m|+1} \|\mathbf{p}_{m,\kappa_{m,l}}-\mathbf{p}_{m,\kappa_{m,l-1}}\|}{V_{max}}.
\end{align}

The goal is to jointly optimize the task assignment strategy, the localization coordinate point and the serving order, and minimize the maximum task completion time among all UAVs, the optimization problem can be formulated as
{
\begin{subequations}
	\label{optimization_problem111}
	\begin{align}
		\label{optimization_problem1}
		P1:&\min_{ \{ \mathcal{G}_m \}, \{ \mathcal{L}_m \}, \{\Gamma_m\} }\max_{m}\left(\frac{\sum_{l=1}^{|\mathcal{G}_m|+1} \|\mathbf{p}_{m,\kappa_{m,l}}-\mathbf{p}_{m,\kappa_{m,l-1}}\|}{V_{max}}\right) \\
		\label{optimization_problem11}
		\text{s.t.}\quad&\bigcup_{m=1}^M \mathcal{G}_m = \mathcal{S}, \\
		\label{optimization_problem12}
		&E_{m} \leq E_{m}^{max}, \forall m,\\
		\label{optimization_problem13}
		&\mathbf{q}_m(t^{e}_m) = \mathbf{q}_m(0), \forall m,\\
		\label{optimization_problem14}
		&e_s(t^e_m) \leq e_{th}, \forall s,\\
		\label{optimization_problem15}
		&|| \mathbf{v}_m(t) || = V_{max} , \forall m, 0 \leq t \leq t^e_m.
	\end{align}
\end{subequations} 
}Denote $E^{max}_m$ and $E_m$ as the total energy and the energy consumption of $m$-th UAV, respectively.
where (\ref{optimization_problem11}) indicates that UAVs need to locate all trapped people. 
(\ref{optimization_problem12}) gives a limit on the maximum total energy consumption of all UAVs.
(\ref{optimization_problem13}) indicates that UAVs will eventually return to the starting point, $t^e_m$ represents the mission completion time for $m$-th UAV. 
(\ref{optimization_problem14}) gives the requirement for localization accuracy.
(\ref{optimization_problem15}) gives a limit on the maximum speed of the UAVs.

\section{Localization-Oriented UAV Adaptive Trajectory Planning}

\subsection{Initial Scan}

\begin{figure}[!t]
	\centering
    \setlength{\abovecaptionskip}{-0.19cm}
	\includegraphics[width=0.311\textwidth,height=0.32\textwidth]{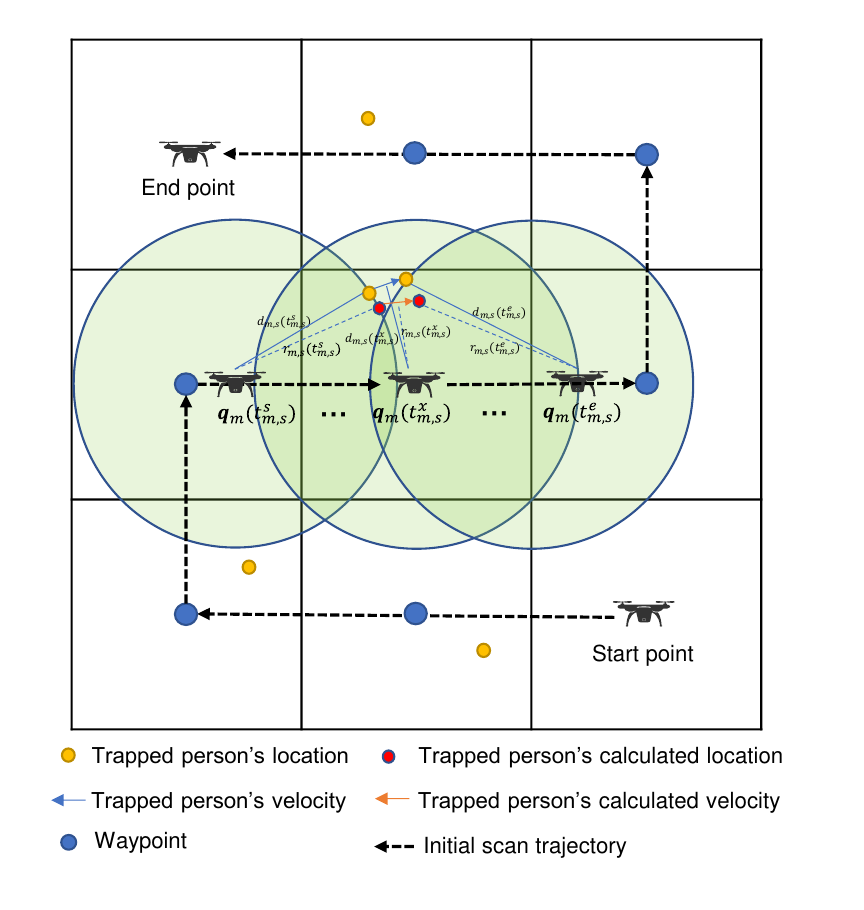}
	\caption{Initial scan trajectory of 1 UAV in the disaster area} 
	\label{fig2}
\end{figure}

In practice, it is difficult to obtain prior information about the number and exact location of trapped people. 
Therefore, the initial scan is employed by UAVs to find the coarse location information of each trapped person.
A reasonable scan trajectory needs to be planned to optimize the scan time and energy consumption cost. 
Specifically, the disaster area is divided into a limited number of equal cells, and when the UAV is located in the center of the cell, its communication range can cover the entire cell. For ease of analysis, we use the center of these cells as the waypoint during the initial scan. It is worth noting that UAV does not hover at the waypoint. The waypoint is just for convenience to represent the trajectory of the UAV's initial scan. 
To facilitate the expansion to other shapes, we assume that the sensors are randomly distributed in a square area in this paper.
Fig. \ref{fig2} shows the initial scan procedure with the number of UAV $M=1$. 
If there are multiple UAVs, the disaster area is divided into large areas equal to the number of UAVs, and these multiple UAVs are each responsible for the initial scan task of one of the large areas.

The purpose of the initial scan is that regardless of the trapped person's location, there will be at least one UAV capable of continuously receiving the trapped person's signal for a certain period, such that enough data can be obtained to compute the velocity of the trapped person. 
Let $t^s_{m,s}$ and $t^e_{m,s}$ denote the start time and end time when the $m$-th UAV receives the RSSI signal from $s$-th trapped person, and $\mathbf{r}_{m,s} = \{r_{m,s}(t_{m,s}^s),...,r_{m,s}(t_{m,s}^e)\} $ and $\mathbf{d}_{m,s} = \{d_{m,s}(t_{m,s}^s),...,d_{m,s}(t_{m,s}^e)\}$ denote the corresponding estimated distance set and actual distance set, respectively.\footnote{As shown in Fig. \ref{fig2}, some areas will be scanned repeatedly, which causes the UAV may receive the signal of the trapped person in two periods of time. Take one of them as an example, to avoid cumulative error, whenever trapped person signals are continuously received on a new route, the continuous signals received in the past are no longer used.} 
According to \eqref{eq6}, the joint PDF of $\mathbf{r}_{m,s}$ can be expressed as

{\small
	\begin{align}
		\label{eq7}
		f(\mathbf{r}_{m,s})=\prod_{t=t_{m,s}^s}^{t_{m,s}^e} \frac{1}{{r_{m,s}(t)\sqrt{2\pi} \sigma}{/\eta}}	\exp \left(\frac{-(\ln (r_{m,s}(t) /  d_{m,s}(t)))^2}{2\sigma^2/\eta^2}\right). 
	\end{align}
}Therefore, the MLE of $\mathbf{d}_{m,s}$ is given by
\begin{equation}
	\label{eq8}
	\mathbf{d}^{*}_{m,s} = \underset{\mathbf{d}_{m,s}}{{\arg\max}} P(\mathbf{r}_{m,s}).
\end{equation}
Converting \eqref{eq8} to logarithmic form with respect to $\mathbf{d}^{*}_{m,s}$ and maximizing it, we have
{\smaller[2]
	\begin{align}
		\label{eq9}
		\mathbf{d}^{*}_{m,s} = \underset{\mathbf{d}_{m,s}}{{\arg\max}} \sum_{t=t_{m,s}^{s}}^{t_{m,s}^e} \left[ -\ln( \frac{r_{m,s}(t)\sqrt{2\pi}\sigma}{\eta}  ) - \frac{(\ln ( r_{m,s}(t) / d_{m,s}(t)) )^2}{2\sigma^2/\eta^2} \right].
	\end{align}
}

In practice, a UAV will collect the signal of the trapped person multiple times in a time slot, and the UAV and the trapped person's location do not change in a time slot.
Since the time slots are very short, it can be assumed that the velocity of the trapped people do not change in several time slots,
according to this setting and \eqref{eq_w_m_t} and \eqref{eq5}, there is enough data to solve \eqref{eq9} to get the sequences  $\{\mathbf{w}_{s}^e(t_{m,s}^s),...,\mathbf{w}_{s}^e(t_{m,s}^e) \}$ and $\{\mathbf{v}_{s}^e(t_{m,s}^s),...,\mathbf{v}_{s}^e(t_{m,s}^e) \}$, where $\mathbf{w}_{s}^e(t)$ and $\mathbf{v}_{s}^e(t)$ are the calculated location and velocity of $s$-th trapped person, respectively.
Let $\mathbf{v}_s^{e,m} =  \frac{\mathbf{w}_{s}^e(t_{m,s}^e) - \mathbf{w}_{s}^e(t_{m,e}^s)}{t_{m,s}^e-t_{m,s}^s}$ as average velocity between time slot $t_{m,s}^s$ and $t_{m,s}^e$.
Therefore, when the UAV cannot receive the trapped person's signal, the trapped person's location in time slot $t$ can be calculated by 
\begin{align}
	\label{eqlo}
	\mathbf{w}_{s}^e(t) =  \mathbf{w}_{s}^e(t_{m,s}^e) + \mathbf{v}_s^{e,m} (t-t_{m,s}^e), t_{m,s}^e  \leq t.
\end{align}

\begin{figure}
	\centering
    \setlength{\abovecaptionskip}{-0.19cm}
	\includegraphics[width=0.33\textwidth,height=0.22\textwidth]{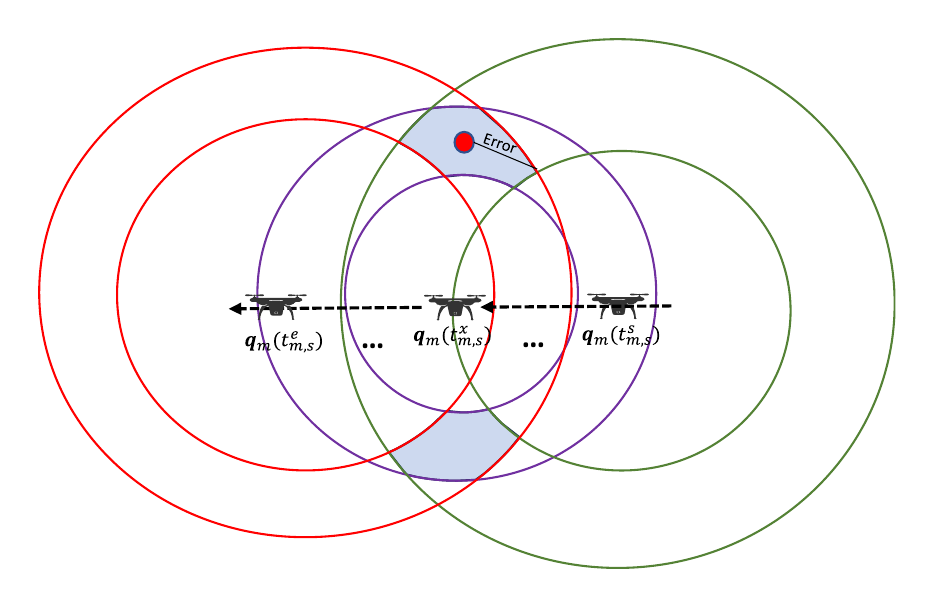}
	\caption{Localization error calculation} 
	\label{fig-define_error}
\end{figure}

Fig. \ref{fig-define_error} shows how we represent the localization accuracy metric in time slot $t_{m,s}^e$, the red dot represents the location of the trapped person in time slot $t_{m,s}^e$ calculated using MLE. 
For $T$ time slots from $t_{m,s}^s$ to $t_{m,s}^e$, generate $T$ sets according to 
{\smaller[1]
	\begin{align}
		&I_{m,t} = d^*_{m,s}(t) - m_{\xi}(d^*_{m,s}(t)) -{v}_{max} (t_{m,s}^e-t),  \\
		&O_{m,t} = d^*_{m,s}(t) + m_{\xi}(d^*_{m,s}(t)) + {v}_{max} (t_{m,s}^e-t),  \\
		\label{define_annulus}
		&\mathcal{A}_{m,t} = \{(x, y)| I_t^2 \leq  (x_m(t)-x)^2+(y_m(t)-y)^2 \leqslant O_t^2 \wedge (x,y) \in  \mathbb{R}^{2\times 1} \}, \
	\end{align}
}where  $\mathcal{A}_{m,t}$ represents the possible locations of trapped person in time slot $t_{m,s}^e$ obtained from the ranging information of different time slots.
The geometric form of the set $\mathcal{A}_{m,t}$ is represented as an annulus, and the overlapping blue areas of multiple annuli represent the possible locations of the trapped people (three points are taken for display in the figure), and the location error $e_s(t_{m,s}^e)$ can be obtained by finding the farthest boundary point from the estimated location to the blue area. According to (2), we consider velocity error as $\alpha e_s(t_{m,s}^e)$, where $\alpha$ is the weight factor determined by prior information. Therefore, the localization error of the $s$-th trapped person in time slot $t$ can be expressed as:
\begin{align}
	\label{eq_error}        
	e_s(t) = e_s(t_{m,s}^e) + \alpha e_s(t_{m,s}^e)(t-t_{m,s}^e),t_{m,s}^e < t.
\end{align}

However, two problems need to be taken into account:

1) For the trapped people, the waypoints where the UAV can receive the RSSI signal may all be on a straight line. In this case,  \eqref{eq_w_m_t}, \eqref{eq5} and \eqref{eq9} can only solve two symmetrically possible locations and velocities along the scan trajectory.

2) According to \eqref{eq_error}, the localization error is positively correlated with the distance. If the distance between the fixed trajectory of the initial scan and the trapped person exceeds a threshold, the localization accuracy cannot meet the needs of search and rescue scenarios.

\subsection{Accurate Localization}

\begin{figure}
	\centering
    \setlength{\abovecaptionskip}{-0.19cm}
	\includegraphics[width=0.3\textwidth,height=0.31\textwidth]{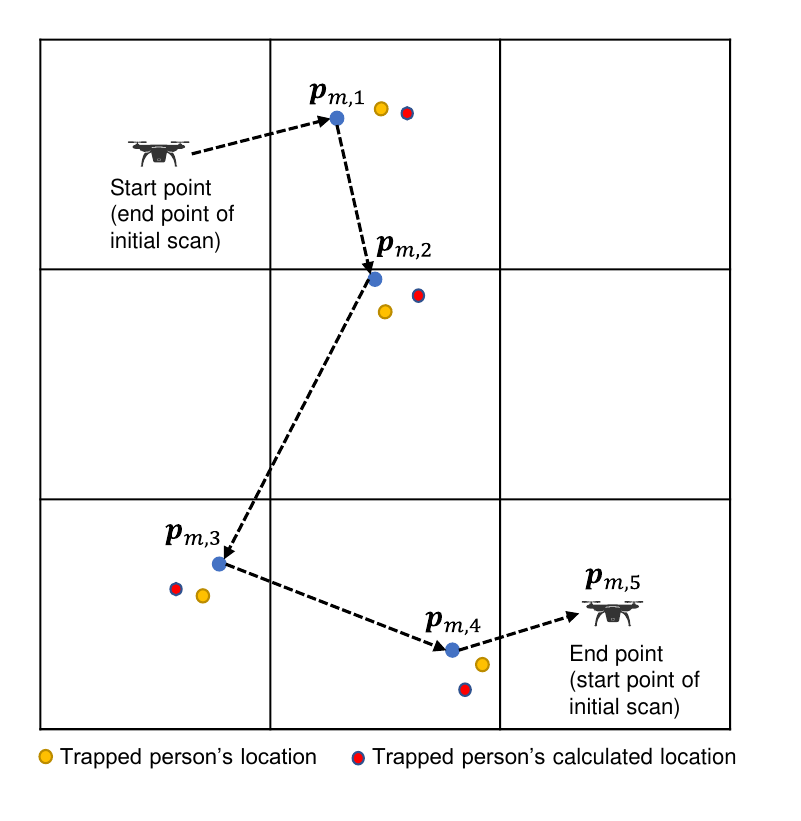}
	\caption{Trajectory of the UAV in the accurate localization stage  } \label{fig3}
\end{figure}

In the accurate localization stage, UAVs will launch from the endpoint of the initial scan, the $m$-th UAV will fly to a location close to the trapped people in $\mathcal{G}_m$ to measure distance and perform location calculation with the same scheme during the initial scan, and then return to the start point of the initial scan. 
As illustrated in Fig. \ref{fig3}, the trajectory of the UAV in the accurate localization stage is a set consisting of a series of broken lines which solves the issue of the initial scan trajectory being a line segment.

$P1$ involves the optimization of the UAV trajectory, which is a continuous function in relation to time $t$, as well as UAV-trapped person association. Therefore, solving $P1$ is challenging due to its infinite number of optimization variables.

To reduce the algorithm's time complexity, we solve $P1$ in two steps. The first step considers a special case that all the $\mathbf{p}_{m,l}$ have been given, e.g., $\mathbf{p}_{m,l} = \mathbf{w}_{\mathcal{G}_m[l]}^e(t_{m,l}^a)$, where $t_{m,l}^a$ and $\mathcal{G}_m[l]$ represent the time slot from $m$-th UAV to $l$-th waypoint and $l$-th elementary in $\mathcal{G}_m$, respectively. 
This indicates that the UAV will fly directly above the location of the trapped person calculated by \eqref{eqlo}. 
This special case is equivalent to the Multiple Traveling Salesman Problem (MTSP), and high-quality approximate solutions can be found efficiently using heuristic algorithms. 
We solve this special case to get $\{\mathcal{G}_m\}$ and  $\{\Gamma_m\}$, such that each trapped person is visited by only one UAV.

Flying above the trapped person for distance measurement may obtain a localization accuracy higher than $e_{th}$. 
To expedite the search and rescue mission, accuracy is sacrificed for quicker task completion.
In the second step, waypoints $\{\mathcal{L}_m\}$ are determined for each UAV to further reduce mission completion time. 
We propose an edge access strategy integrated into the iterative solution process of the first step to obtain a high-quality approximate solution to $P1$ under normal circumstances.

The strategy is as follows: 
we define the observation location as the center of a circle with a radius $e_{s}(t_{m,s}^a)$. Generate $R$ reference points on the circle denoted as $\mathcal{U}_n=\{\mathbf{u}_{1},...,\mathbf{u}_{R}\}$, where  $t_{m,s}^a$ as the time slot when $m$-th UAV reaches one of the waypoints in $\{\mathcal{L}_m\}$. 
The selected waypoints should make the localization accuracy of all reference points greater than $e_{th}$. According to \eqref{eq_error}, we can create another circle with the observation location as the center. As long as the waypoints are located in the circle, the localization accuracy requirements can be met. 
When UAVs fly to the edge of the circle, they fly directly to the next waypoint, then the flight length and flight time of the UAVs will be reduced.

PSO can be used to solve the MTSP problem in the first step. However, traditional PSO still has problems such as slow convergence and prone to local optimality.
Given the above shortcomings, an enhanced PSO (EPSO) algorithm is proposed.
And combine the edge access strategy and EPSO's fitness function to solve $P1$.
The specific design is as follows:

$\mathbf{Position}$: The UAV-trapped person association $\{ \mathcal{G}_m \}$ and the serving order $\{\Gamma_m\}$, denoted as $x^{k}_i$. Each particle will update its position according to the update rule:
\begin{align}
	\label{eq_position}
	x^{k+1}_i = x^{k}_i + v^{k+1}_i.
\end{align}

$\mathbf{Velocity}$: Each particle updates its velocity $v^k_i$ by learning the particle's personal best position $pb_i^k$ and the global best position $gb^k$,  the  update rule is :
{ \smaller[1]
	\begin{align}
		\label{eq_velocity}
		v^{k+1}_i= \epsilon × v^k_i+c_1 × rand(0,1) × (pb_i^k - x^k_i)+ c_2 × rand(0,1) ×  (gb^k - x^k_i),
	\end{align}
}where $\epsilon$ is the inertia factor, $c_1$ and $c_2$ are the individual optimal and global optimal weight parameters, respectively.
To speed up the convergence and avoid local optimum, we use some tricks to optimize the above PSO algorithm.
To balance the global search and local search capabilities, we use the adaptive weight method, the expression of the weight is
{
\begin{align}
	\label{eq_epsilong}
	\epsilon=\left\{
	\begin{aligned}
		 &  \epsilon_{min} + \frac{(\epsilon_{max}-\epsilon_{min})(f-f_{min})}{f_{avg}-f_{min}} &, f \leq f_{avg}\\
		 &   \epsilon_{max} &, f > f_{avg}
	\end{aligned}
	\right.
\end{align}
}
Among them, $\epsilon_{min}$ represents the minimum value of the inertia factor; $\epsilon_{max}$ represents the maximum value of the inertia factor; $f$ represents the temporary fitness; $f_{min}$ represents the minimum value of the temporary fitness; $f_{avg}$ represents the average value of the temporary fitness. For particles whose objective function value is better than the average target value, the corresponding inertia weight factor is smaller, thus retaining the particle. On the contrary, for particles whose objective function value is worse than the average target value, the corresponding inertia weight factor is larger, so that the particle moves closer to the better search area.

$\mathbf{Fitness}$ $\mathbf{Value}$: Based on $\{ \mathcal{G}_m \}$,  $\{\Gamma_m\}$ of the particle and the given $p_{m,l}$, use the edge access strategy to optimize $p_{m,l}$ and use \eqref{eqtm} to calculate the  fitness function.

For the location of the $l$-th trapped person, if the initial scan obtains two possible solutions, we randomly select a location as $p_{m,l}$. 
If the UAV approaches a certain waypoint and finds another possible location that is closer to the real location. The UAV flies to the location for localization. During this flight, the trajectory planning algorithm is run again until the accurate localization of all trapped people is completed. 
The entire algorithm including the initial scan is called MLE-EPSO, and EPSO is described in detail in Algorithm 1.

{
\begin{algorithm}[t] 	
\small
	\DontPrintSemicolon
	\SetAlgoLined
	\KwIn {The initial scan obtains the trapped people's coarse location and coarse velocity, the localization error, and UAVs' current location}
	\KwOut { $\{ \mathcal{G}_m \}$, $\{ \mathcal{L}_m \}$, $\{\Gamma_m\}$, the trapped people's precise location and precise velocity}
	\While{UAVs not fully traversed $\{\mathcal{L}_m\}$ }{
		$\{\mathcal{L}_m\} \leftarrow \varnothing$ \;
		\For{trapped people not be accurately located}{
			\eIf{initail scan get a certain location or accurate location excluded the error location}{
				join $\{\mathcal{L}_m\}$
			}{
				Randomly choose a solution to join $\{\mathcal{L}_m\}$
			}
		}
		Random initialization particle swarm on $\{\mathcal{L}_m\}$\;
	 	\For{$k = 1$ to iteration}{
	 		Select the highest fitness value particle as $gb_k$.\;
	 		\For{ i = 1 to population }{
                    Calculate the fitness function using edge access strategy and \eqref{eqtm} to update $pb^k_i$.\;
	 			Update the velocity $v^k_i$ and position $x^k_i$  based on \eqref{eq_position}, \eqref{eq_velocity} and \eqref{eq_epsilong}, respectively.\;
 			}
 		}
 		\SetKwRepeat{Do}{do}{while}	
 		\Do{It was found that the solution in the initial scan stage closer to the true location of the trapped person was not added to $\{\mathcal{L}_m\}$}{
 			UAVs traverse $\{\mathcal{L}_m\}$ on $\{\Gamma_m\}$
 		}
	}
	\caption{Enhanced PSO Algorithm}

\end{algorithm}

}

\section{Simulation And Performance Analysis}

In this section, we present simulation results to evaluate the performance of the proposed method.
We simulate a rectangular area of 1120$m$ × 640$m$ to send 4 UAVs. The UAVs fly at a height of 100$m$, communicate within a distance of 150$m$, and travel at a velocity of 25$m/s$, trapped people cannot move faster than 1.5$m/s$,  each time slot is set to 0.025$s$, and the UAV samples RSSI signal 10 times in each time slot.
For the odometry model, the parameter $\eta = 2$ and $\sigma_\Psi = 4$dB.
For UAV power consumption, we adopt the parameter settings from \cite{7888557-1}.

\begin{figure*}
	\centering
	\subfigure[Localization task completion time with different numbers of trapped people]{
		\begin{minipage}[b]{0.315\textwidth}
			\includegraphics[width=1\textwidth,height=0.7\textwidth]{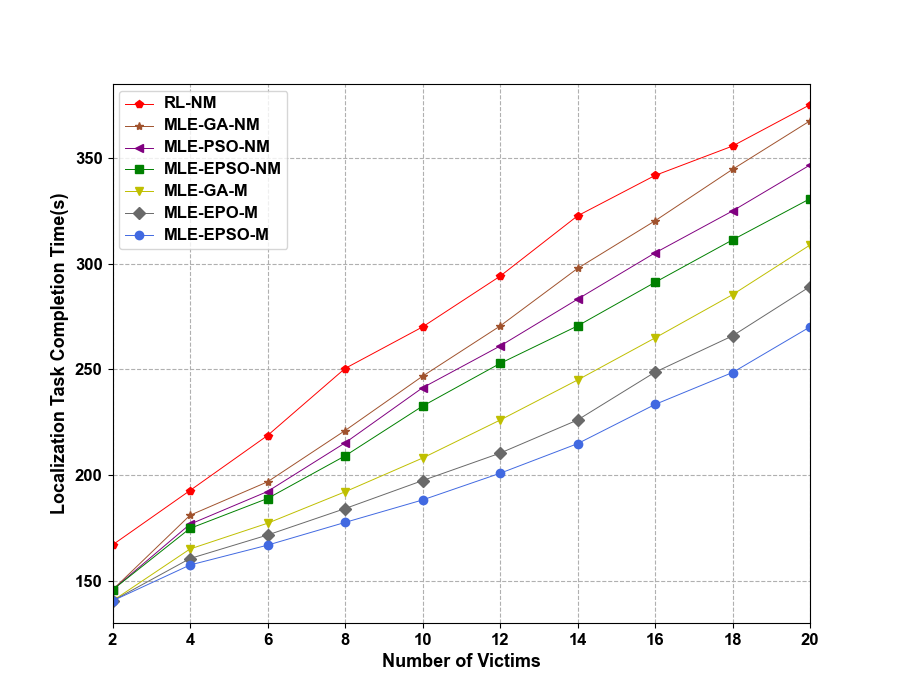}
		\end{minipage}
		\label{simulation_1}
	}
	\subfigure[Algorithm training run time with different numbers of trapped people]{
		\begin{minipage}[b]{0.315\textwidth}
			\includegraphics[width=1\textwidth,height=0.7\textwidth]{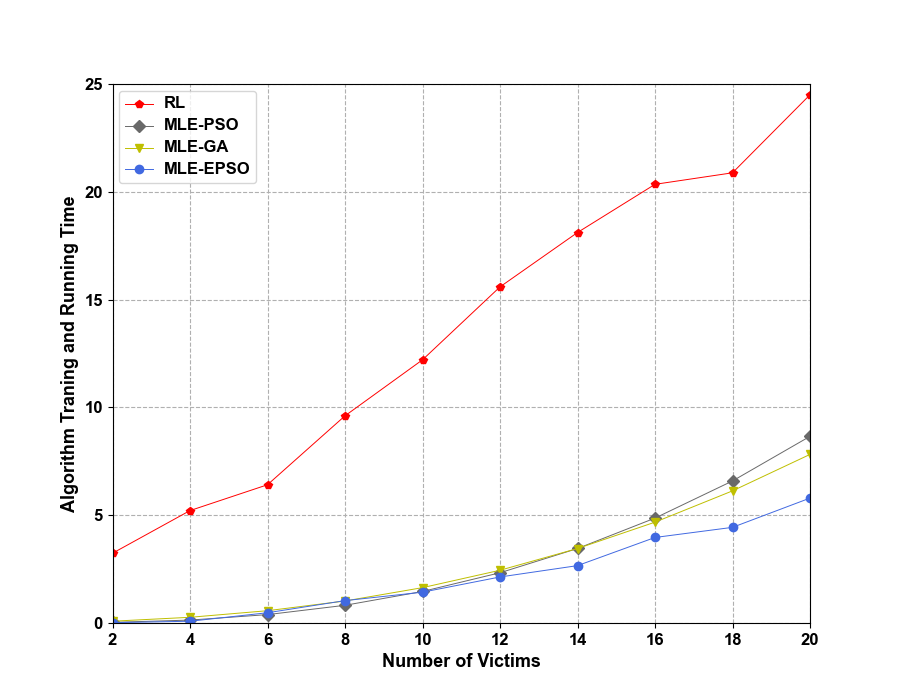}
		\end{minipage}
		\label{simulation_2}
	}
	\subfigure[Maximum localization error of different task completion time when the number of trapped people = 10]{
		\begin{minipage}[b]{0.315\textwidth}
			\includegraphics[width=1\textwidth,height=0.7\textwidth]{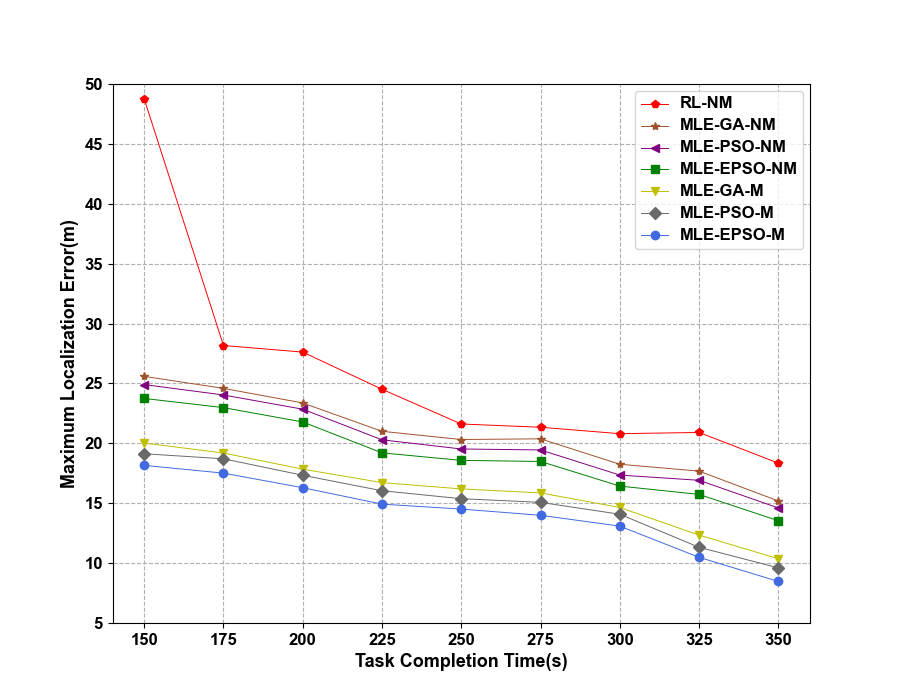}
		\end{minipage}
		\label{simulation_3}
	}
	\caption{Performance comparison in terms of task completion time, algorithm runtime and localization error}
	\label{fig:hor_2figs_1cap_2subcap}
\end{figure*}

Comparative solutions, such as \cite{8960453}, do not utilize initial ranging information to determine the location of the trapped people. Instead, they employ the RL algorithm to address the optimization problem $P1$ directly. In our study, we modify the rewards and optimization goals to align with \eqref{optimization_problem1} to suit our specific scenario. This approach, referred to as RL-NM, disregards the movement of trapped people, denoted as NM (No Move). Conversely, the scheme denoted as M (Move) takes into account their movement.
This comparison scheme aims to verify the improvement of the overall algorithm using the single anchor point positioning strategy based on the MLE proposed by us.
Additionally, the EPSO component of MLE-EPSO utilizes PSO and genetic algorithm (GA) as comparative solutions respectively to assess the performance of EPSO.

In Fig. \ref{simulation_1}, we compare the time required to achieve a given localization accuracy.
Among them, the methods using MLE are better than RL-NM. The reason is that RL-NM uses a hovering strategy and $p_{m,l}$ discretization design, which will increase the task completion time and may lose the optimal solution.
The MLE-based single anchor point localization method adopts the UAV non-hovering strategy and does not require discrete resolution of space.
The method that considers user movement is better than the method that does not consider user movement. The reason is that the method that does not consider user movement needs to achieve higher localization accuracy in the accurate localization stage.
To meet the localization accuracy requirements when localizing the target. This will undoubtedly increase flight distance and mission completion time.
For the three algorithms EPSO, PSO, and GA, EPSO has achieved the best results. The reason is that the algorithm we proposed will retain better particles and let the worse particles quickly move closer to the better search area.

In Fig. \ref{simulation_2}, we compare the running time of different algorithms. 
Considering the distribution and number of trapped people in emergency search and rescue scenarios, the RL method requires retraining in new scenarios, so the compared time includes the training time of RL.
No matter how the number of trapped people changes, the MLE-EPSO/PSO/GA method has a shorter algorithm running time. As the number of trapped people increases, the state space of the MLE-EPSO/PSO/GA method also increases, and the algorithm running time also increases. Increase. The RL-NM method always has a larger state space. Although the state space does not change with the number of trapped people, the increase in the number of trapped people will make the environment more complex and the algorithm convergence time will increase.

In Fig. \ref{simulation_3}, we compare the localization accuracy achieved by different algorithms for a given task completion time. The more time a given task is completed, the smaller the localization error can be achieved. The RL-NM adopts $p_{m,l}$ discretization design, which may lose the optimal solution. We use the MLE method in the initial scan to reduce the state space and ensure localization accuracy without discretization.

\section{Conclusions}

This paper investigates the task of UAV adaptive trajectory planning for dynamic user localization in emergency search and rescue scenarios. 
In terms of localization, the MLE method is used to fully mine the ranging information, and the non-hovering strategy is integrated to achieve single-anchor localization, to solve the problem of limited available UAVs in the early stage of disaster relief.
Additionally, an estimation of the mobile user's location is obtained, taking user movement into account and providing an upper bound on the localization error.
Combining this localization method and localization error-index, in terms of trajectory planning,
the localization task is divided into two stages: initial scan and accurate localization. 
The initial scan stage aims to acquire approximate trapped person locations as prior information for subsequent algorithms.
In the accurate localization stage, to reduce the algorithm's time complexity, the problem is decomposed and the EPSO algorithm and edge access strategy are employed for localization-oriented adaptive trajectory planning. This approach accelerates algorithm convergence and further reduces task completion time.
Simulation results show that this scheme reduces the time required to complete the localization task by 28.7\% compared to the comparative scheme.
Furthermore, it achieves higher localization accuracy within the same task duration, while maintaining lower time complexity for the algorithm.
\ifCLASSOPTIONcaptionsoff
\newpage
\fi

\bibliographystyle{IEEEtran}
\bibliography{references.bib}

\end{document}